\documentclass[,final]{aipproc}
\layoutstyle{6x9}

\begin{document}

\title{GRB 070714B - Discovery of the Highest Spectroscopically Confirmed Short Burst Redshift}


\author{J. F. Graham}{
  address={Space Science Telescope Institute}
  ,altaddress={Johns Hopkins University}}
\author{A. S. Fruchter}{address={Space Science Telescope Institute}}
\author{A. J. Levan}{address={University of Warwick}}
\author{M. Nysewander}{address={Space Science Telescope Institute}}
\author{N. R. Tanvir}{address={University of Leicester}}
\author{T. Dahlen}{address={Space Science Telescope Institute}}
\author{D. Bersier}{address={Liverpool John Moores University}}
\author{A. Pe'er}{address={Space Science Telescope Institute}}

\begin{abstract}
Gemini Nod \& Shuffle spectroscopy on the host of the short GRB 070714B shows a single emission line at 7167 {\AA} which, based on a grizJHK  photometric redshift, we conclude is the 3727 {\AA} [O II] line.  This places the host at a redshift of z=.923 exceeding the previous record for the highest spectroscopically confirmed short burst redshift of z=.546 held by GRB 051221.  This dramatically moves back the time at which we know short bursts were being formed, and suggests that the present evidence for an old progenitor population may be observationally biased.
\end{abstract}

\maketitle

\subsection{Introduction}

GRB 070714B was initially detected by the Burst Alert Telescope (BAT) on the NASA SWIFT spacecraft.  The gamma ray emission consisted of several short spikes with a collective duration of 3 seconds followed approximately twenty seconds later by fifty seconds of softer emission.  The main component also shows a small spectral lag (Norris et al. GCN 6631 \cite{GCN6529}).    This emission is similar to previous short bursts including GRB 050724 and places this burst securely in the short category. 

Rapid follow observations with the 2m Liverpool Telescope and the 4m William Herschel Telescope found a fading optical transient within the XRT error circle.  Subsequent observations detected a host galaxy at the location of the optical transient.

Here we report on photometry and spectroscopy of the host galaxy, which lead us to conclude that the host is a moderately star-forming galaxy at a redshift of z=0.923.  The spectroscopy and redshift discussed here was originally reported in GCN 6836 (Graham et al. 2007 \cite{GCN6836}).

\subsection{Host Galaxy Imaging}

Optical imaging was obtained in g, r, i, and z bands with the GMOS instrument on the Gemini North Telescope.  Due to a narrow widow between the object rising and dawn, observations were preformed in non-photometric conditions shortly before and often extending slightly past astronomical twilight, at airmasses just below and sometimes slightly above 2, and had to be spread across several nights.  Photometric calibration observations were subsequently preformed via observations in photometric conditions of the object field and selected area 95 of Landolt (1992) \cite{Landolt1992}.  Science and calibration observations were scaled from field star fluxes.  Absoulte flux was determined from SDSS magnitudes of the Landolt stars.  Magnitudes are listed in table \ref{mags}.

Near-infrared imaging was obtained in J, H and K with the NIRI instrument on Gemini North on the nights of July 24$^{th}$, 25$^{th}$ and 26$^{th}$, 2007.  Single exposures were taken totaling one minute each in coadds of 5 x 12s, 6 x 10s and 5 x 12s, in J, H and K respectively.  The images were taken under photometric conditions and low airmass.

\begin{ltxtable}[h]\center
\begin{tabular}{|c|c|}
\hline
band    &   mag  $\pm$ error \\
\hline
g    &   25.79 $\pm$ .34   \\
r    &   24.90 $\pm$ .21   \\
i    &   23.97 $\pm$ .12   \\
z    &   24.01 $\pm$ .13   \\
J    &   22.27 $\pm$ .12   \\
H    &   22.28 $\pm$ .20   \\
K    &   21.13 $\pm$ .13   \\
\hline
\end{tabular}
\caption{Photometric magnitudes of the host galaxy in each observed band.\label{mags}}
\end{ltxtable}

\subsection{Host Galaxy Spectroscopy}

Initial spectroscopic observations were obtained on the night of the July 25$^{th}$.  Due to the drop in detector sensitivity long-ward of 9250 {\AA} a central wavelength of 7250 {\AA} was selected yielding a spectral range of 5250 to 9250 {\AA}.  Further spectroscopy was conducted on the night of September 13$^{th}$ with the central wavelength shifted out to 7750 {\AA}.  The R400 grating offerers a reasonable compromise between spectral resolution and width of coverage and was used both nights.  Due to the abundance of skylines in the spectral range the Nod \& Shuffle method was used.  The first and second night of spectroscopy consisted of four and six 10-minute Nod \& Shuffle exposures respectively.

The individual spectroscopic exposures were reduced using the standard Nod \& Shuffle process. This resulted in eight and twelve 300 second images for the 7250 and 7750 {\AA} central wavelengths respectively.  To optimize cosmic ray rejection the positive and negative (once inverted) images of each central wavelength were combined in a single step.  The two central wavelengths were then combined weighted by total exposure time.  Spectral extraction was preformed with IRAF task appall using a 10 pixel wide aperture in the spatial direction.  The continuum was too weak to allow automatic tracing of the aperture center.  However a tracing of a bright star also present in the slit showed insignificant variation in the spatial location of the continuum along the spectra negating the need for automatic tracing.  A fixed center aperture was thus used.

This yielded a spectrum with a spectral resolution of 1.37 {\AA} per pixel and a spectral resolution of 0.15 arc seconds per pixel.  A single spectral line was observed at 7167 {\AA} with a measured equivalent width of $-42.89 \pm 4.09$ and flux of $2.0\pm.3*10^{-17}$ erg/s/cm$^{2}$.  Aside from a faint continuum no other spectral features were detected in the 5150 to 9900 {\AA} spectral range.  

\subsection{Spectroscopic Line Determination}

The optical and IR photometry also allows us to constrain the 7167 {\AA} line identity via a photometric redshift determination.  There are three reasonable candidates for the observed spectral line at 7167 {\AA}; the 3727 {\AA} [O II] line placing the object at a redshift of .92, the 5007 {\AA} [O III] line placing the object at a redshift of .43, and the 6563 H$\alpha$ line placing the object at a redshift of .09.  

\begin{figure}[h!]
\includegraphics[width=.45\textwidth]{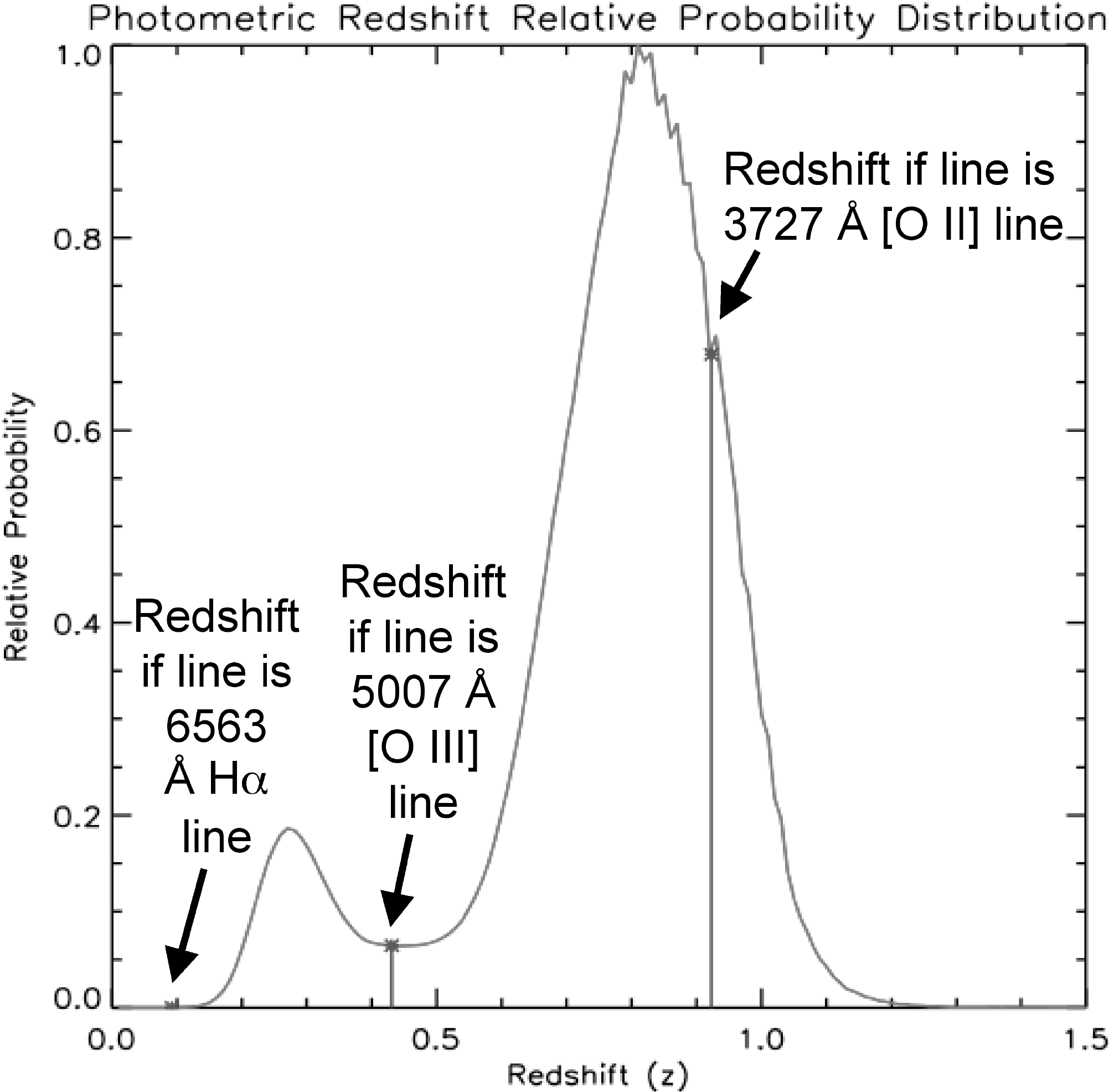}
\caption{Photometric redshift relative probability distribution with the various line possibilities annotated.  \label{prob}}
\end{figure}

These possibilities along with the best-fitting photometric redshift (and their respective best-fitting spectral type for the host galaxy) are calculated using the template-fitting method. In this method, the observed photometry is matched to synthetic photometry derived using the filter through-puts for the instrument used and a set of library templates redshifted in the range $0<z<6$. The spectral templates used here consists of the E, Sbc, Scd, and Im templates from Coleman et al. (1980) \cite{Coleman} , together with two starburst templates from Kinney et al. (1996) \cite{Kinney}. The calculated a photometric redshift probability distribution for the host galaxy is show in figure \ref{prob}.% and table \ref{probs}.

\begin{figure}[ht]
\includegraphics[width=.5\textwidth]{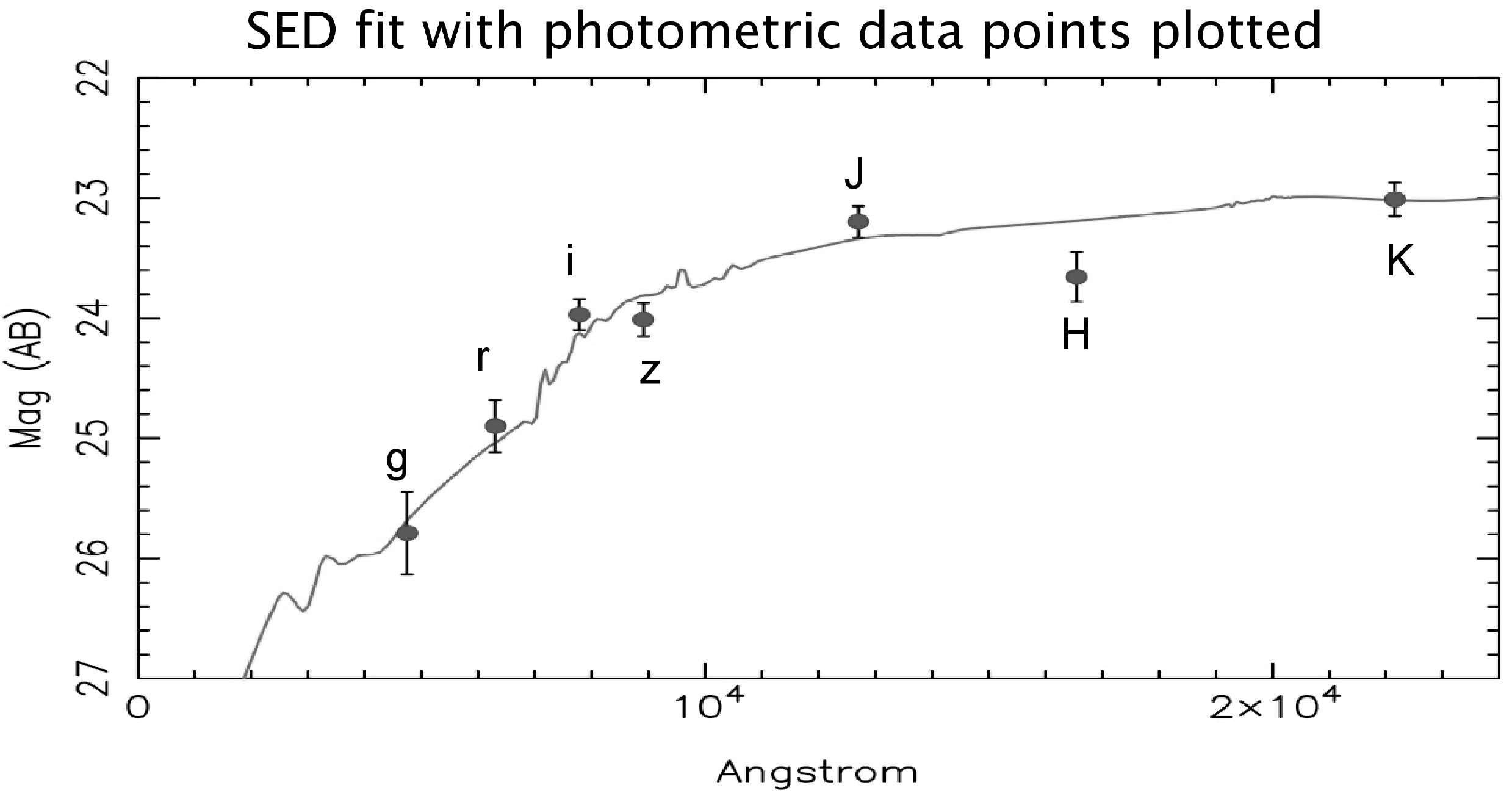}
\caption{The Spectral Energy Distribution of the z=.923 fit (a -19.6 V magnitude Scd galaxy) with the measured photometric values over plotted. \label{sed}}
\end{figure}

The detected line is by a factor of ten most likely the 3727 {\AA} [O II] line making the photometric redshift consistent with a GRB host galaxy at $z=0.923$.  The best fit spectral energy distribution (-19.6 V magnitude Scd galaxy) with the measured photometry over plotted is shown in figure \ref{sed}.

\subsection{Conclusions}

Spectroscopy of the host of GRB 070714B reveals a single line at 7167 {\AA}.   Our multiband optical and infrared photometry strongly implies that this can only be the  3727 {\AA} [O II] line at a redshift of z=0.923.  This nearly doubles the highest spectroscopically redshift of a short burst with a sub-arsecond position.   A number of bursts have shown no obvious host, thus suggesting that their hosts might be distant and faint (Berger 2007 \cite{Berger2007})  Additionally, other authors have determined the redshifts of galaxies in XRT error circles, and some of these galaxies have had redshifts also around $z \sim 1$ \cite{GCN5470}\cite{Berger}\cite{GCN5965}\cite{GCN3801}.  However, we note that a fairly small XRT error circle (for a short  burst) with radius of 5" would have a greater than $60\%$ chance of  containing a random galaxy as bright as the host of GRB 070714B.  In cases such as this one where the burst has an optical afterglow, the host, and thus the redshift, can be determined with very little chance of confusion.  This observation thus moves back the time at which we know short bursts were being formed to the first half of the age of the universe, and suggests that the present evidence for an old progenitor population may be, at least partially, observationally biased.

\bibliographystyle{aipproc}
\bibliography{\jobname}

\end{document}